\documentclass[%
 aip,
 amsmath,amssymb,
 reprint,%
]{revtex4-1}

\usepackage{graphicx}
\usepackage{dcolumn}
\usepackage{bm}
\usepackage[utf8]{inputenc}
\usepackage[T1]{fontenc}
\usepackage{mathptmx}
\usepackage{xcolor}
\usepackage{textcomp, gensymb}
\usepackage{hyperref}
\hypersetup{
    colorlinks=true,
    linkcolor=blue,
    filecolor=magenta,
    urlcolor=cyan,
}
\begin{document}


\title{Characteristic signatures of Northern Hemisphere blocking events in a Lagrangian flow network representation of the atmospheric circulation}

\author{No\'{e}mie Ehstand}
\email{n.ehstand@ifisc.uib-csic.es}
 \affiliation{IFISC  (CSIC-UIB), Instituto  de  F\'{\i}sica  Interdisciplinar y Sistemas Complejos,
 Campus Universitat de les Illes Balears, E-07122 Palma de Mallorca, Spain}

\author{Reik V. Donner}%
\affiliation{
Department of Water, Environment, Construction and Safety, Magdeburg-Stendal
University of Applied Sciences, Breitscheidstra{\ss}e 2, 39114 Magdeburg, Germany
}
\affiliation{
Research Department IV -- Complexity Science and Research Department I -- Earth System Analysis,
Potsdam Institute for Climate Impact Research (PIK) -- Member of the Leibniz Association,
Telegrafenberg A31, 14473 Potsdam, Germany
}%

\author{Crist\'{o}bal L\'{o}pez}
\affiliation{%
IFISC  (CSIC-UIB), Instituto  de  F\'{\i}sica  Interdisciplinar y Sistemas Complejos,
Campus Universitat de les Illes Balears, E-07122 Palma de Mallorca, Spain
}%

\author{Emilio Hern\'{a}ndez-Garc\'{\i}a}
\affiliation{%
IFISC  (CSIC-UIB), Instituto  de  F\'{\i}sica  Interdisciplinar y Sistemas Complejos, Campus Universitat
de les Illes Balears, E-07122 Palma de Mallorca, Spain
}%

\date{\today}

\begin{abstract}
In the past decades, boreal summers have been characterized by an increasing number of extreme weather events in the Northern Hemisphere extratropics, including persistent heat waves, droughts and heavy rainfall events with significant social, economic and environmental impacts. Many of these events have been associated with the presence of anomalous large-scale atmospheric circulation patterns, in particular persistent blocking situations, i.e., nearly stationary spatial patterns of air pressure. To contribute to a better understanding of the emergence and dynamical properties of such situations, we construct complex networks representing the atmospheric circulation based on Lagrangian trajectory data of passive tracers advected within the atmospheric flow. For these \emph{Lagrangian flow networks}, we study the spatial patterns of selected node properties prior to, during and after different atmospheric blocking events in Northern Hemisphere summer. We highlight the specific network characteristics associated with the sequence of strong blocking episodes over Europe during summer 2010 as an illustrative example. Our results demonstrate the ability of the node degree, entropy and harmonic closeness centrality based on outgoing links to trace important spatio-temporal characteristics of atmospheric blocking events. In particular, all three measures capture the effective separation of the stationary pressure cell forming the blocking high from the normal westerly flow and the deviation of the main atmospheric currents around it.
Our results suggest the utility of further exploiting the Lagrangian flow network approach to atmospheric circulation in future targeted diagnostic and prognostic studies.
\end{abstract}

\maketitle

\begin{quotation}
As the frequency and severity of mid-latitude extreme weather events such as heat waves, droughts and heavy rainfall events is projected to further increase with ongoing climate change, developing reliable forecasts of such events is becoming a gradually more pressing issue. However, while the quality of predictions has been improving considerably on short-term (up to 2 weeks lead time) and seasonal time scales (beyond 3 months), sub-seasonal forecasting (from 2 weeks to about 3 months) remains a challenging task. This results in part from a limited understanding and representation of phenomena that could potentially increase the predictability at these sub-seasonal scales. One type of such phenomena are atmospheric blocking events. These large-scale, nearly stationary, atmospheric pressure patterns can remain in place for several days or even weeks, disturbing the usual westerly driven circulation and resulting succession of weather regimes over the mid-latitudes. Despite numerous studies, a comprehensive theory explaining the emergence of blocking-related circulation anomalies and allowing an early forecasting of incipient blocking situations remains to be found. In this work, we utilize a network-based approach, so-called Lagrangian flow networks, for studying the atmospheric circulation associated with blocking situations during Northern hemisphere summer. We discuss the ability of different network measures to detect and track important spatial characteristics of blocking events, suggesting the potential of complex network approaches to provide key elements for future diagnostic and prognostic studies of atmospheric blocking events.
\end{quotation}

\section{\label{sec:introduction}Introduction}

The study of transport and mixing in time-dependent flows is
central in many areas of science ranging from astrophysics to
biological dynamics, climate, plasma physics or engineering.
Important research efforts have been devoted to understanding
the fundamental mechanisms by which fluid movement is
organized, enhanced, and limited, as well as the ways by which it might be
controlled. Various techniques have been developed in this
context. In particular, Lagrangian dynamics based approaches using
concepts from dynamical systems theory have undergone important
developments in the past
decades \cite{Mancho2006,Lacasce2008,Samelson2013,Haller2015, vanSebille2018}.
These approaches can
be roughly  classified into two main categories, depending on the
mathematical tools being used: geometric or probabilistic. The
first class exploits stretching fields and related quantities to identify transport barriers in the flow,
associated with Lagrangian coherent structures
\cite{Haller2015,Haller2000,dOvidio2004,LEKIEN20051,
Bettencourt2013, HALLER2000352}. The second category builds upon the transfer
operator of the system, which describes the evolution of
densities of particles passively advected within the flow. The
properties of the discretized transfer operator
\cite{Froyland2010,Froyland2015,Bolltbook2013}, understood as a
transition matrix, allow us to directly detect coherent fluid
regions rather than the transport barriers between them. In several previous studies~\cite{SerGiacomi2015,Rossi2014,Bolltbook2013, Ser-GiacomiPRE, Donner2019, Iacobello2021, RodriguezMendez2017}, the authors proposed an interpretation of
the transition matrix as the adjacency matrix of a weighted and
directed network, referred to as a Lagrangian flow network, thereby
allowing the utilization of powerful tools and metrics developed in the
context of graph theory to explore the flow properties. These
different approaches are naturally connected and combining them
can lead to a deeper understanding of the flow dynamics
\cite{FROYLAND20091507,Bolltbook2013,SerGiacomi2015,Lindner}.

There exist many examples of successful applications of the probabilistic methods mentioned above in geophysical contexts. For example, Froyland \emph{et~al.}~\cite{Froyland2015}
used a methodology based on transfer operators to study the pathways of oceanic structures known as Agulhas rings. Rossi \emph{et~al.}~\cite{Rossi2014} and Dubois \emph{et~al.}~\cite{Dubois2016} used a Lagrangian flow network approach to characterize marine connectivity in the Mediterranean Sea, allowing for a better understanding of the spatial organization of marine populations and the design of marine protected areas. A similar connectivity analysis was recently performed in the Arctic \cite{Reijnders2021}. Other related studies analysed the surface connectivity in the Mediterranean derived from drifter data \cite{celentano2020} and studied the transition paths of marine debris \cite{Miron2021}.

Applying the geometric or probabilistic methodologies to the atmosphere is even more challenging than in the ocean. In fact, while the ocean surface is almost divergence free at large scales and can hence be treated as a $2$-dimensional flow, the vertical component of the atmospheric circulation is more noticeable, especially in the tropics due to the importance of convective phenomena. This poses computational as well as visualization challenges. An additional and perhaps more fundamental difficulty stems from the general lack of separation between the time scales of material advection in the atmosphere and the time scales of motion and deformation of the structures of interest. There have been, however, several successful studies using both approaches. Examples include studies of the Antarctic polar vortex \cite{Koh2002,Joseph2002,delaCamara2012,Santitissadeekorn2010,Froyland2010}, of atmospheric rivers \cite{GaraboaPaz2015}, or the identification of humidity sources \cite{MartinGomez2016}.

In this work, we apply a Lagrangian flow network approach to investigate the atmospheric circulation dynamics associated with blocking events. Blocking highs are large scale, nearly stationary (persisting for at least 4 days up to several weeks) atmospheric pressure patterns occurring in the mid to high latitudes. They are known to be associated with an anomalously strong and quasi-stationary meandering of the subpolar jet stream and are often linked with extreme weather events~\cite{Woollings2018, Martius2012, Kornhuber2017}. One of the most prominent examples of the past decades occurred in the summer of 2010. During the months of June, July and August, several successive anomalously persistent blocking anticyclones appeared over Eastern Europe, preventing storms and frontal systems from reaching the region, hence resulting in a strong heatwave which lasted for several weeks over Western Russia. Furthermore, the persistence of the blocking highs allowed for cold air masses to be channeled southwards along their downstream edge towards the North of Pakistan and India where they interacted with moist tropical air masses provided by the Indian summer monsoon circulation. The combined effect of the extratropical disturbance, monsoon surges and the mountainous topography of the area, further enhanced by the presence of La Ni\~na conditions during that year, led to extreme rainfalls and floods in the North of Pakistan~\cite{Hong2011, Martius2012, Lau2012}. Similar situations have occurred repeatedly in the past decades with significant social, economic and environmental repercussions. Yet, despite numerous studies, the characterization and understanding of the mechanisms behind the dynamics of these types of circulation anomalies as well as their connections with other atmospheric events still poses open challenges to future research.

Recently, Ser-Giacomi \emph{et al.}~\cite{SerGiacomi2015-blocking} demonstrated that optimal paths in a Lagrangian flow network highlight distinctive circulation patterns associated with atmospheric blocking events. Building upon this work, in this paper we construct sequences of Lagrangian flow networks covering the three boreal summers of 2003, 2010 and 2018, all of which have been characterized by multiple strong blocking situations. For this purpose, network nodes are associated with different regions of the atmosphere, and links are established based on the flux of material between these nodes during a given time interval. Given that blocking highs are long-lived coherent structures with a persistence of days to weeks, and acknowledging the recent successes of the network approach as detailed above, the question we will address is the following: How well do Lagrangian flow network measures reflect such nearly stationary atmospheric circulation patterns? In particular, we focus on three network measures: the degree, the entropy and the harmonic closeness centrality of each node when accounting for outgoing links only. We demonstrate the ability of these measures to trace the spatio-temporal characteristics of blocking events and in particular the perturbation of the subpolar jet stream related to the presence of blocking highs.

This paper is organized as follows. In Section~\ref{sec:network-description}, we summarize the Lagrangian flow network formalism and define the selected network measures to be investigated during and outside of atmospheric blocking situations. Section~\ref{sec:computational-aspects} describes the data sources used in our work and the computational aspects related to the construction of the network. Finally, our results are presented in Section~\ref{sec:results} before concluding in Section~\ref{sec:conclusion}.

\section{\label{sec:network-description} Network description of Lagrangian Transport in Fluid Flows}

\subsection{Lagrangian dynamics: flow field and flow map}
We consider the Lagrangian motion of passive tracers in a
time-dependent velocity field $\mathbf{v}(\mathbf{x},t)$
defined at the points $\mathbf{x}$ of the finite domain $\Omega
\subset \mathbb{R}^3$, which in our case will be a portion of
the Earth's atmosphere. The tracer trajectories are obtained by
integrating the ordinary differential equation
\begin{equation}
    \frac{d\mathbf{x}}{dt} = \mathbf{v}(\mathbf{x},t),
    \label{eq:flow}
\end{equation}
with suitable initial conditions. The flow map
$\Phi(\mathbf{x}_0,t_0;\tau):~\Omega \times \mathbb{R} \times
\mathbb{R} \rightarrow \Omega$ associated to the above system
gives the final position of a tracer released at $\mathbf{x}_0$
at time $t_0$ and evolved under the dynamics of the flow for a
fixed time $\tau$:
\begin{equation}
    \mathbf{x}(t_0 + \tau) = \Phi(\mathbf{x}_0, t_0; \tau).
\end{equation}

For simplicity, in the following we use the brief notation
$\Phi(\mathbf{x}, t_0; \tau) =\mathrel{\mathop:}
\Phi_{t_0}^\tau (\mathbf{x})$. Further, note that when applied
to a whole set $A$ the flow map specifies the movement of the
entire fluid region: $A(t_0+\tau) = \Phi_{t_0}^\tau(A)$.

\subsection{Numerical approximation of the dynamics -- the discrete transfer operator}
In order to obtain a spatially discretized version of the
Lagrangian flow dynamics described above, we first divide
$\Omega$ into a set of $N$ mutually disjoint boxes $\{B_i\}_{i=1}^{N}$ such that
$\Omega = \bigcup B_i$. For the time interval $[t_0,
t_0+\tau]$, the proportion of mass displaced from $B_i$ to $B_j$
under the action of the flow defined by Eq.~\eqref{eq:flow} is given by
\begin{equation}
    P(t_0, \tau)_{ij} \mathrel{\mathop:}=
    \frac{{m}(B_i \cap \Phi_{t_0 + \tau}^{-\tau}(B_j) )}{{m}(B_i)}\ \ \ \forall\ i,j\ \in\ \{1,\ldots,N\},
    \label{eq:transfer_operator}
\end{equation}
where $m$ is a normalized volume measure over $\Omega$.
The elements $P_{ij}$ define a transfer operator
$\mathbf{P}(t_0, \tau)$ encoding the conditional transition
probability between any two boxes. Note that $\mathbf{P}$ is a
discrete version of the Perron-Frobenius operator obtained
within the framework of Ulam's approach~\cite{Ulam1960, Koltai2010}. If the
flow is closed within $\Omega$, $\mathbf{P}$ is row stochastic.

In order to approximate numerically the entries defined in
Eq.~\eqref{eq:transfer_operator}, we need to estimate the amount of
fluid exchanged between each pair of boxes in the interval
$[t_0, t_0+\tau]$. To do so, a large number $M_i$ of tracers are
initialized in each box $B_i$ and advected for a time $\tau$ in
the velocity field, allowing us to compute
\begin{equation}
    P_{ij} \approx \frac{\#\{\mathbf{x}:\mathbf{x}(t_0)\in B_i\  \textnormal{and}\  \mathbf{x}(t_0+\tau)\in B_j\}}{M_i}.
    \label{eq:weights}
\end{equation}
For the purpose of the present work, in the following
each box $B_i$ will represent a vertical column
spanning the full height of the atmosphere and characterized by a
horizontal area in circular coordinates. Hence, although
exploiting the fully three-dimensional dynamics of the trajectories
$\mathbf{x}(t)$, we will focus only on the analysis of the horizontal
transport between these columns. On practical grounds, we can imagine $\{B_i\}_{i=1}^{N}$ as two-dimensional
boxes tessellating a part of the surface of the Earth, with
$P_{ij}$ describing horizontal transport among them.
In addition, we will always consider the case in
which all boxes $B_i$ have the same horizontal size. As a result, since the number of particles initialized in each box accounts for the amount of fluid contained in that box, we will also take the initial numbers of particles $M_i$ to be equal in each box: $M_i\equiv M\ \forall i \in \{1,\ldots,N\}$.

\subsection{Network properties}
Following previous work~\cite{SerGiacomi2015,Bolltbook2013}, the transfer operator $\mathbf{P}(t_0,
\tau)$ defined in Eq.~\eqref{eq:transfer_operator} can be
interpreted as the adjacency matrix of a temporal, weighted and
directed network encoding the transport process -- the Lagrangian
flow network. Precisely, the network consists of $N$ nodes
representing the $N$ boxes $\{B_1, B_2,\ldots,B_N\}$. A link
exists between node $i$ and $j$ if $P_{ij}>0$, i.e. if material
is exchanged between boxes $B_i$ and $B_j$, and the weight of the
link is given by the transition probability $P_{ij}$.

A great variety of tools developed in the context of graph and complex network theory ~\cite{newman2010} can be used to study the topology of the Lagrangian flow network and eventually provide new insights into the dynamics of the flow. Among those, our present study focuses on three network measures as presented below.

\subsubsection{Degree} \label{subsec:degree-theory}
Perhaps the simplest centrality measure in a complex network is the degree, measuring how many nodes are connected to each vertex. In a directed network, one must distinguish the \textit{in-degree}, $K_i^{in}$, and \textit{out-degree}, $K_i^{out}$, which in the case of a Lagrangian flow network reflect how many nodes can reach and can be reached from $i$, respectively, within the considered time step $\tau$. In the following, we will study only the characteristics of outgoing links, which are related to the dispersal of material in the studied flow. The corresponding out-degree is computed from the adjacency matrix as
\begin{equation}
    K_i^{out} = \sum_{j} \Theta(P_{ij})\quad i,j\ \in \ i=1,\ldots,N,
    \label{eq:degree}
\end{equation}
where $\Theta(\bullet)$ is the Heavyside function with $\Theta(x)= 0 $ for $x\leq 0$ and $\Theta(x) = 1$ for $x > 0$. Note that this definition allows for self-connections, that is, it also includes trajectories that end within $i$ after completion of the time step. For a fixed cell size $\Delta^2$, the total area covered by the nodes that received fluid from node $i$ within the given time step is approximately equal to $K_i^{out}\Delta^2$. Thus, intuitively, the degree can be related to the "stretching" of the flow at $i$. In fact, Ser-Giacomi \emph{et al.}~\cite{SerGiacomi2015} showed heuristically that in regions dominated by hyperbolic structures, the degree is related to the cell average of the maximum eigenvalue of the Cauchy-Green strain tensor associated to the flow, $C(\mathbf{x}_0, t_0, \tau) = (\nabla \Phi_{t_0}^\tau(\mathbf{x}_0))^T\nabla \Phi_{t_0}^\tau(\mathbf{x}_0)$, via \[ K_i^{out} \approx \langle ~ \sqrt{\Lambda} ~ \rangle_{B_i}, \] where $\Lambda$ is the maximum eigenvalue of $C(\mathbf{x}_0, t_0, \tau)$.

\subsubsection{Entropy}
The degree weighs all nodes equally, independently of the amount of matter they receive, and therefore tends to exhibit high values in regions where dispersion is particularly important. To correct for the corresponding effect, it can be beneficial to account for the heterogeneity in the fluxes sent from node $i$ to all the other nodes, which can be conveniently included in the associated node \emph{entropy}.

Again, we focus on the corresponding quantity characterizing the outgoing links, which can be defined as~\cite{SerGiacomi2015}
\begin{equation}
     H_i^{out}(t_0, \tau) = -\frac{1}{\tau}\sum_j P(t_0, \tau)_{ij}\textnormal{log}(P(t_0, \tau)_{ij}).
\end{equation}
This definition corresponds to the set-oriented finite-time entropy previously studied by Froyland and Padberg-Gehle \cite{FROYLAND20121612}. It is a probabilistic measure of the uncertainty about the final position of particles initialized in cell $i$. In flows dominated by hyperbolic structures, and for sufficiently large integration times, the entropy approximates the value of the finite-time Lyapunov exponent averaged over each box \cite{SerGiacomi2015}, thus highlighting again the stretching experienced by the flow in cell $i$. This connection is related to standard results on the relation between the Kolmogorov-Sinai entropy and the sum of the positive Lyapunov exponents as presented by Boffetta \emph{et al.}~\cite{Boffetta_2002}.

\subsubsection{Harmonic closeness centrality} \label{subsec:closeness-theory}
Both the degree and the entropy are local measures that take only information on the direct neighbors of a given node into account. A non-local measure of centrality is the (weighted) \textit{harmonic out-closeness centrality} of a node.

In general, the harmonic closeness centrality measures the mean of the inverse graph distances from a given node to all other nodes and is thus defined as
\begin{equation}
     C_i = \frac{1}{N-1}\sum_{j\ne i}\frac{1}{d(i,j)},
\end{equation}
where $d(i,j)$ is the graph distance along the shortest path going from node $i$ to node $j$. More precisely, let this path be $\mathcal{I}=\{i,n_1, n_2, \ldots , n_{m-1}, j\}$. Then $d(i,j)=l_{in_1} + l_{n_1n_2} + \ldots +l_{n_{m-1}j}$, where $l_{n_\mu n_\nu}$ are the individual link distances between nodes. In our particular case, we are interested in the connectivity of nodes in terms of material flux and thus define the individual link distances as $l_{n_\mu n_\nu} = ({P_{n_\mu n_\nu}})^{-1} \ \ \forall \ P_{n_\mu n_\nu}\ne 0$. Defined so, the closeness naturally gives more importance to nodes which belong to "chains" of cells connected via large flux exchange.

In the context of atmospheric circulation, it is important to recall that the flow (and, hence, its network representation) is time dependent. Thus, the individual links defining $d(i,j)$ only reflect the topological state of the network characterizing the flow dynamics in the interval $[t, t+\tau]$, but do not represent true particle trajectories beyond that time. Nonetheless, assuming that the flow pattern changes sufficiently slow, i.e. a certain coherence in the currents between subsequent intervals, the following can be expected. First, due to the "chain" property of the closeness, the nodes within the main flow currents should have higher closeness values than their surroundings. Second, since the out-closeness value at a given node depends on the connection state of the nodes located downstream of it, any perturbation of flow currents should cause the closeness values to gradually decrease upstream of the perturbation.

\section{\label{sec:computational-aspects} The atmospheric Lagrangian Flow Network: computational aspects}

\subsection{Description of the data} \label{subsec:data}
In this study, we use data from the European Centre for Medium-Range Weather Forecasts (ECMWF) ERA-Interim reanalysis product \cite{Dee2011}. ERA-Interim is a global atmospheric data set, covering a 40-year period from January 1979 to August 2019. It includes information at 60 vertical pressure levels (from the surface up to $0.1$~hPa) with a spatial resolution of $0.75^\circ\times0.75^\circ$ in latitude and longitude and a time step of $6$~hours.

The variables provided include dew point temperature,
geopotential height, land cover, planetary boundary layer
height, air pressure and pressure reduced to mean sea level (MSL),
relative humidity, temperature, zonal and meridional component
of the wind, vertical velocity and water equivalent to
accumulated snow depth, which are all necessary to run the
particle integration model Flexpart described next.

\subsection{Flexpart -- a flexible particle dispersion model}
\label{subsec:Flexpart}
As mentioned in Section~\ref{sec:network-description}, the construction of the Lagrangian flow network is based on the trajectories of tracers evolving under the dynamics of the flow. In order to obtain the required data, we use here the Flexpart model version 10.3 \cite{Pisso2019} which simulates particle transport and mixing in the atmosphere by integrating Eq.~\eqref{eq:flow}.

More precisely, the software has to be provided with the necessary atmospheric data (in our case obtained from the ERA-Interim data set) including not only the zonal, meridional and vertical components of the wind field, but also additional meteorological fields such as temperature, pressure, humidity, etc. All fields are then interpolated by Flexpart at the simulated particles' positions and the mean flow transport $\mathbf{\bar{v}}$ is complemented with stochastic components $\mathbf{v^t}$ and $\mathbf{v^m}$ to better represent turbulent and mesoscale wind fluctuations occurring at small scales, so that the velocity field on the right hand side of Eq.~\eqref{eq:flow} becomes $\mathbf{v}=\mathbf{\bar{v}} + \mathbf{v^t} + \mathbf{v^m}$. The turbulent components in $\mathbf{v^t}$ are parameterized assuming a Markov process based on the Langevin equation with parameters depending on the local meteorological state. The mesoscale fluctuations in $\mathbf{v^m}$ are obtained by solving an independent Langevin equation, assuming that the variances of the wind at the grid scale and sub-grid scale are related. For additional details, we refer to Stohl \emph{et al.}~\cite{Stohl2015}.

In practice, Flexpart allows the user to specify the number of particles to be simulated, their release altitude and horizontal position, as well as the release time and duration of integration. The output of the software contains the three-dimensional position of all particles along with additional information such as the potential vorticity, humidity, temperature and air density at each point along the trajectory.

\subsection{Domain discretization and trajectories integration}
\label{subsec:domain-and-integration}
In the present work, we focus on the Northern Hemisphere atmospheric circulation. The computational domain, however, extends into the Southern Hemisphere in order to avoid undesired effects due to loss of mass at the domain boundaries resulting from cross-equatorial mass transport. More precisely, we consider the domain $180\degree$W-$180\degree$E / $25\degree$S-$85\degree$N, discretized into 7272 two-dimensional cells of equal area using a sinusoidal projection. Each cell is defined by two longitudinal and two latitudinal coordinates, ($\phi_0$,$\phi_1$) and ($\theta_0$,$\theta_1$), and its area on the surface of the Earth $\mathcal{A}$ is given by $\mathcal{A} = R^2(\phi_1-\phi_0)(\textnormal{cos}\theta_0-\textnormal{cos}\theta_1)$, where $R = 6371$~km is the Earth's radius.

\begin{figure}
\centering
\includegraphics[scale =1.1]{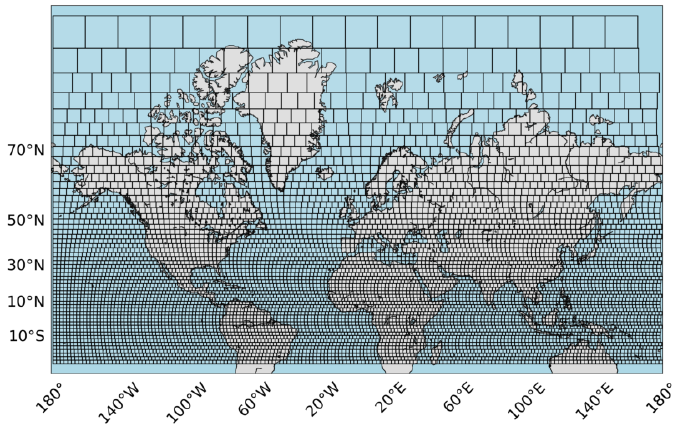}
\caption{Domain discretization defining the flow network's nodes.}
\label{fig:grid}
\end{figure}

For the purpose of the present study, we choose as the common area that of a $2\degree \times 2\degree$ cell at the equator, that is, $\mathcal{A} \approx 49447~\textnormal{km}^2$. The resulting grid is shown in Fig.~\ref{fig:grid}. Note that since we do not allow cells to overlap and require them to have equal area, the domain cannot be entirely covered. Hence, a small section of it is excluded at the Eastern boundary. However, this section is practically negligible and does not affect our results qualitatively. Furthermore, note that the grid size is much coarser than the resolution of the ERA-Interim data used to integrate the trajectories (see Section~\ref{subsec:data}). Thus, small scale features appearing in the trajectories are not explicitly described by the network but are included in a statistical way.

In each grid cell, we initialize $M$ tracers (see below) at a pressure level of $300$~hPa, which are then advected in the three-dimensional wind field for a time $\tau$ using Flexpart (see Section~\ref{subsec:Flexpart}). Although the obtained trajectories are also three-dimensional, the final positions are projected onto the longitude-latitude plane before computing the components of the transfer operator $P_{ij}$ \eqref{eq:transfer_operator}. We thus ignore any height difference between the initial and final states, based on the assumption that the motion in the mid to high latitudes (where atmospheric blocking patterns are expected to occur) is sufficiently well approximated as two-dimensional. This was indeed verified to be the case for the horizontal scales considered here, that is, scales of the order of the blocking structures sizes ($\approx 1000$ km).

The approximation of two-dimensionality appears justified in the mid-latitudes because of the geostrophic balance of the flow. Such an approximation would not be appropriate in situations where convective activity dominates, as is typically the case in the tropics for instance. In such situations, our methodology, designed to characterize the horizontal connectivity of the flow, would not be able to distinguish between structures present at different heights and a three-dimensional flow network construction would be necessary.

To complete this description, the integration time $\tau$ and the number $M$ of simulated tracers must be specified. The integration time is chosen as $\tau=1$~day for the degree and entropy and $\tau=8$~hours for the harmonic closeness. These values have been selected after comparing the results for different values of the integration time for representative flow configurations. Specifically, we started with choosing an initial value $\tilde{\tau}$ of the same order as the time-scale of coherence of blocking structures and then iteratively refined this choice to obtain prominent network structures that allow for a possible interpretation (see below). The number of released tracers $M$ had to be set sufficiently high to capture all significant parts of the dynamics even in cells with high dispersion. We thus required $M$ to satisfy $K_i^{out} < \alpha M \ \forall i$ with $\alpha = 0.1$. This leads to the choice of $M=900$ for integration time $\tau =8$~hours and $M=1500$ for $\tau = 1$~day, respectively.

\begin{figure*}
\includegraphics{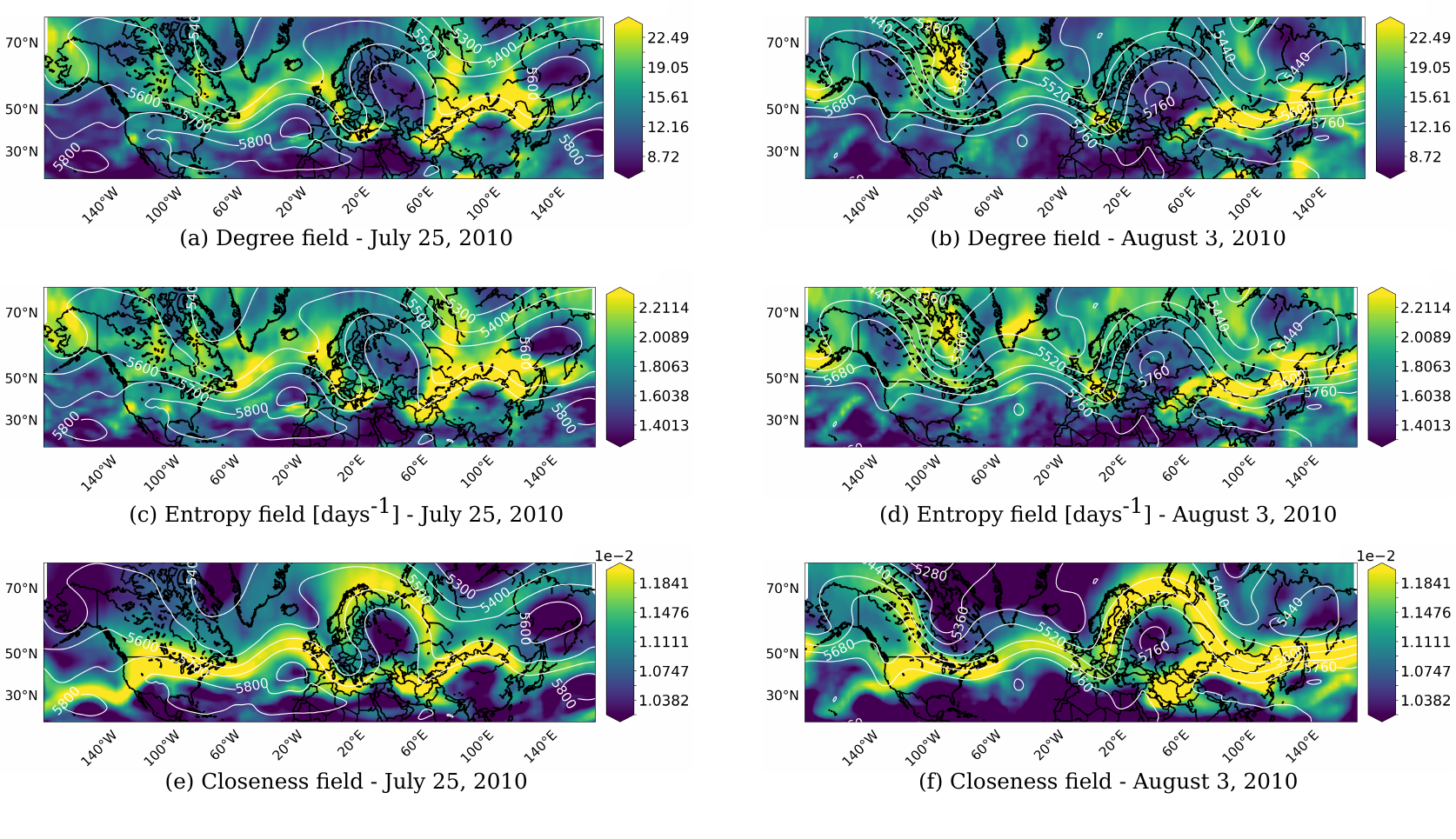}
\caption{Degree, entropy and harmonic closeness centrality fields (color) for two representative time intervals during the sequence of summer 2010 blocking events. Geopotential height contours at 500 hPa are shown as reference (white lines). All fields have been averaged over 4 days from 25 July 2010 to 29 July 2010 (left column) and 3 August 2010 to 7 August 2010 (right column), respectively.}
\label{fig:dec-blocking-2010}
\end{figure*}

\section{\label{sec:results} Results}

In order to identify objectively blocking situations, we computed the blocking index TM90 as described in the seminal paper of Tibaldi and Molteni~\cite{TIBALDI1990} for the boreal summer months June, July and August of three different years with strong heatwaves (2003, 2010 and 2018) focusing on Eurasia. The employed detection methodology for blocking events is based on identifying the meridional reversal of the 500~hPa geopotential height gradient around a constant latitude, 60$^\circ$N, which is considered representative of the climatological jet stream. This determines local and instantaneous blocking conditions. Additional criteria are then imposed to ensure that the detected events have a minimum spatial extension of $15^\circ$ in longitude and persistence of 5 days. Note that there exist, beyond the definition considered here, a large variety of alternative methodologies accounting for different features of blocking events. These methodologies mostly differ in the field considered (geopotential height, potential vorticity, etc.), the vertical levels chosen and the sophistication of the computation itself. In particular, modifications of the classical TM90 methodology have been developed to account for effects such as spatial or seasonal variations of the jet stream~\cite{ANewPerspectiveonBlocking,Barnes2011}. The TM90 index has also been extended to 2D by applying the same ideas to a latitude band instead of one specific latitude only~\cite{Davini2012}. For more general reviews on blocking detection schemes, we refer to refs.~\cite{Woollings2018,Barriopedro2010}.

The events detected using the TM90 index are listed in Table~\ref{table:blockingTM90}. Note that when using different approaches for defining atmospheric blocking or modifying the parameters characterizing the minimum spatial and temporal extent of the respective anomalies, the list of events might be prone to certain changes. In the following, we will focus on the sequence of strong blocking events in summer 2010, which can be considered robust features highlighted by different definitions of atmospheric blocking. Similar results as those presented below have been observed for the 2003 and 2018 blocking events as well. The results for the latter two years can be found in the Supplementary Material. 

\begin{table}[ht]
\begin{center}
\begin{tabular}{ |c|c|c| }
\hline
 Period & Duration & Position \\
 & (days) & (longitude) \\
 \hline
 8-13 June 2003 & $6$ & $106^{\circ}$E \\
 14-19 July 2003 & $6$ & $25^{\circ}$E\\
 17-22 July 2003 & $6$ & $112^{\circ}$E\\
 26 July - 3 August 2003 & $9$ & $31^{\circ}$E\\
 4-9 August 2003 & $6$ & $171^\circ$E \\
 24-28 August 2003 & $5$ & $29^{\circ}$W\\
 \hline
 22-26 June 2010 & $5$ & $32^{\circ}$E\\
 13-20 July 2010 & $8$ & $26^{\circ}$E\\
 23 July - 9 August 2010 & $18$ & $43^{\circ}$E\\
 \hline
 4-8 June 2018 & $5$ & $112^\circ$E\\
 11-17 June 2018 & $7$ & $167^\circ$E \\
 16 July - 2 August 2018 & $18$ & $25^{\circ}$E\\
 \hline
\end{tabular}
\end{center}
\caption{Northern Hemisphere blocking events identified by the TM90 index in the boreal summers of 2003, 2010 and 2018.}
\label{table:blockingTM90}
\end{table}

\subsection{Degree}
Figures~\ref{fig:dec-blocking-2010}(a,b) show the degree field on two different dates during the last of the three 2010 blocking episodes. Note that the degree field is computed at $0000$, $0600$, $1200$ and $1800$ UTC for each day of the simulation period. In order to highlight more persistent patterns, the displayed fields have been obtained by averaging over $4$ days, precisely from July 25 to July 29 and August 3 to August 7, respectively. Further, the averaged degree field, defined in each box of the network, has been interpolated linearly before plotting to avoid visual artifacts due to the heterogeneous linear size of the individual cells. The geopotential height contours at 500~hPa have been added to the figures for reference. The deformation of the contours around $40^\circ$E indicates the location of the blocking.

We observe that the blocking location coincides with an extended region of low degree, surrounded by regions of higher degree, suggesting that the blocking region has fewer flow connections to the rest of the atmosphere than its surroundings. This confirms that, at least on short time scales (that is, of the order of one day), the circulation within the nearly stationary high pressure cell forming the blocking is relatively isolated from stronger atmospheric currents. Note that similar observations can be made for all events listed in Table~\ref{table:blockingTM90}, though the prominence of the low degree regions themselves depends on the structure of the blocking. In
particular, events consisting of several small pressure cells
as, for instance, the one lasting from July 17 to 22, 2003 (see Supplementary Material), are not equally well identified as those consisting of one single pressure cell.

\begin{figure*}
\includegraphics{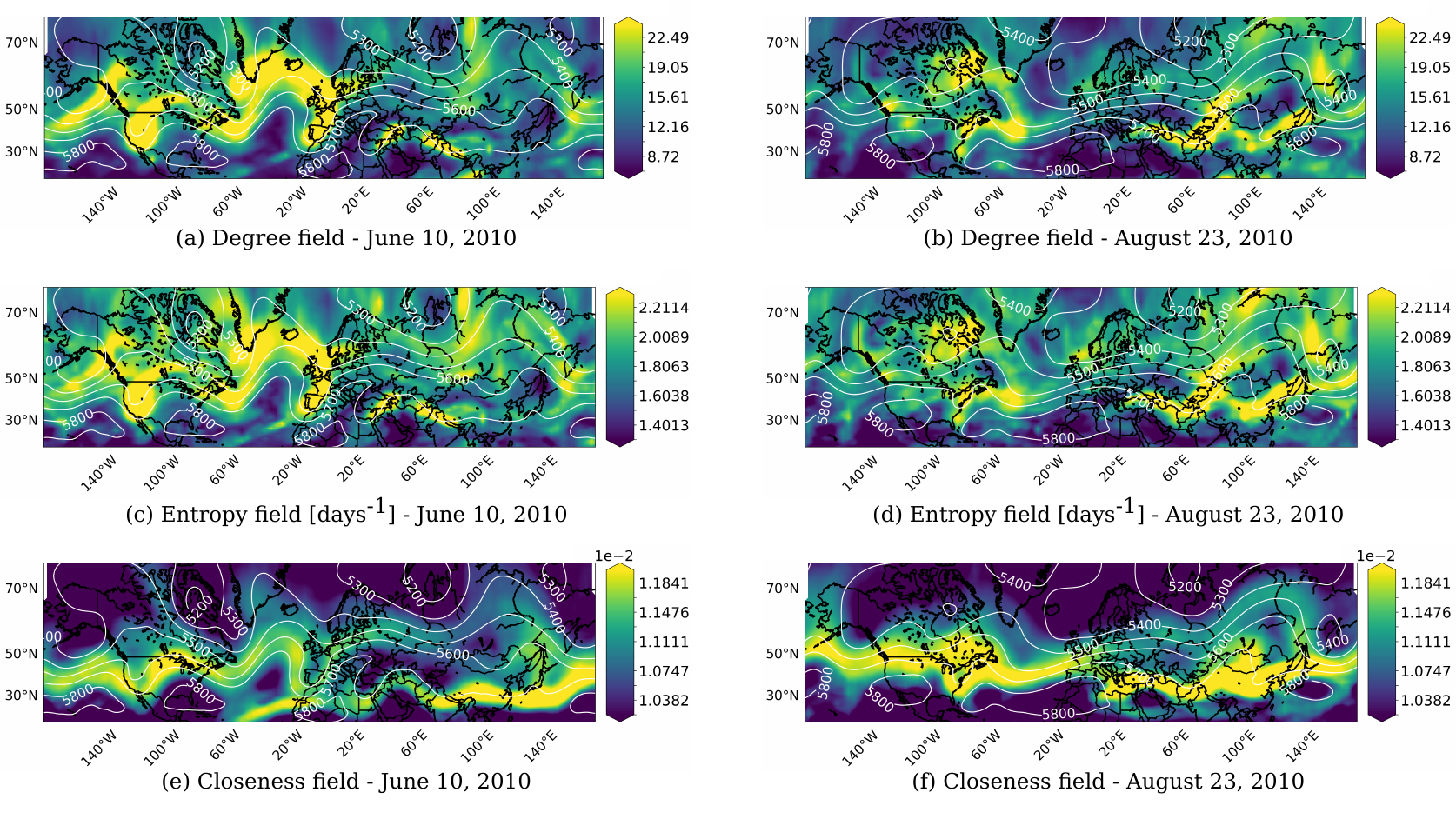}
\caption{Degree, entropy and harmonic closeness centrality fields (color) before and after the sequence of 2010 summer blocking events. Geopotential height contours at 500 hPa are shown as reference. All fields are averaged over 4 days from 10 June 2010 to 14 June 2010 (left column) and 23 August 2010 to 27 August 2010 (right column), respectively.}
\label{fig:dec-outside-blocking-2010}
\end{figure*}

Looking at typical degree fields before and after the sequence of the three 2010 summer blocking periods as exemplified in Figs.~\ref{fig:dec-outside-blocking-2010}(a,b) for periods starting on June 10 and August 23, 2010, one observes the absence of similar extended regions of low degree in the mid-latitudes, confirming further the relation between such regions and the appearance of atmospheric blocking events. Note, however, that small regions of low degree still do exist in the mid-latitudes outside blocking periods, for instance, on June 10, 2010 around $40^\circ$W, $20^\circ$E and $120^\circ$E, and on August 23, 2010 around $50^\circ$E and $120^\circ$E. In fact, any area having relatively little flow exchange with the rest of the study domain on the time scale of the simulated transport is associated with low degree values. It is the spatial extent, latitude and persistence of such region which defines the blocking situation. Hence, while almost every blocking event identified via the TM90 index~\cite{TIBALDI1990} is associated with a region of low degree, the reverse is not necessarily true.

On the other hand, in all figures we can observe streaks of large degree values. As mentioned in Section~\ref{subsec:degree-theory}, these structures highlight regions in which material is significantly stretched under the action of the flow, i.e., regions where the trajectories tend to diverge the most. In general, this is the case at interfaces between different circulation patterns and upstream of perturbations. Precisely, Figs.~\ref{fig:dec-outside-blocking-2010}(a,b) clearly show that the troughs of the geopotential height and the boundaries of pressure cells tend to be associated with high degree values. Similar observations can be made when looking at the troughs east and west of the blocking high perturbation in Figs.~\ref{fig:dec-blocking-2010}(a,b). In addition, the regions where the winds are the strongest naturally tend to be associated with higher degree values because tracers can be advected faster and, hence, dispersed over a larger spatial area. A clear feature appearing in both Figs.~\ref{fig:dec-blocking-2010}(a,b) is a band of large degree values located southeast of the blocking high. The development of this band indeed coincides with a strong jet acceleration downstream of the block between July 23 and August 9, 2010, as already reported in previous works~\cite{Lau2012}. We further notice that on July 25, 2010, the presence of an additional high pressure perturbation downstream of the East Asian jet axis leads to degree values within the jet that are slightly higher than on August 3, 2010.

\subsection{Entropy}
Figures~\ref{fig:dec-outside-blocking-2010}(c), \ref{fig:dec-blocking-2010}(c), \ref{fig:dec-blocking-2010}(d) and
\ref{fig:dec-outside-blocking-2010}(d) present the entropy field before, during and after the 2010 blocking events. As for the degree, the entropy fields are averaged over $4$
days in order to highlight the most persistent structures and interpolated to avoid visual artifacts due to the heterogeneous linear size of the individual cells. The deformation of the geopotential height contours (white lines) around $40^\circ$E indicates the location of the
blocking in Figs.~\ref{fig:dec-blocking-2010}(c,d).

For the sake of simplicity, the node entropy can be thought of as a weighted version of the degree, where the weights, defined by Eq.~\eqref{eq:weights}, represent the transition probabilities between the different network cells. In other words, the links associated with high flow of particles are given more importance as compared to those associated with low flow. Hence, we may expect that the regions and patterns highlighted by the entropy are at least partially similar to those shown by the degree. Indeed, one observes in Figs.~\ref{fig:dec-blocking-2010}(c,d) that the lowest entropy values are found at the blocking location, confirming the relative isolation of the blocking region from the rest of the circulation in the studied cases.
In addition, the troughs and contours of pressure cells are clearly associated with high entropy values in Figs.~\ref{fig:dec-blocking-2010}(c,d) (during the blocking event) and Figs.~\ref{fig:dec-outside-blocking-2010}(c,d) (before/after the blocking). In particular, during the blocking period, high entropy values are also observed surrounding the blocking and in the region of the East Asian jet located southeast of the blocking high (Figs.~\ref{fig:dec-blocking-2010}(c,d)).

Besides these reported similarities, there are however some qualitative differences between the degree and the entropy field. While the degree tends to over-emphasize the regions of high dispersion, the entropy, being a weighted measure, provides a more balanced visualization of the flow stretching in different regions. In this regard, the entropy provides an accurate picture of the relative importance of detected flow anomalies. As an example, in Figs.~\ref{fig:dec-blocking-2010}(a,b) as well as Figs.~\ref{fig:dec-outside-blocking-2010}(a,b), high degree values are found over the Bay of Bengal. These are not accompanied by similarly elevated entropy values, see Figs.~\ref{fig:dec-blocking-2010}(c,d) and \ref{fig:dec-outside-blocking-2010}(c,d). This likely indicates that the high degree observed in this region is related to a high dispersion rather than an important stretching of the flow. In fact, the observed structure might be an artifact due to the important convective motion associated with the monsoon season in this region. Note that, as mentioned in Section~\ref{subsec:domain-and-integration}, the global flow motion is largely horizontal; hence, such effects associated to vertical motion are relatively rare and appear in a very localized manner, commonly close to the equator.

\subsection{Harmonic closeness centrality}\label{subsec:closeness-results}
Finally, we study the harmonic closeness centrality field before, during and after the 2010 blocking periods. While both the degree and the entropy characterize local properties of the network, the closeness is a path-based measure and thus encodes information about global connection patterns in the network.

The corresponding results are shown in Figs.~\ref{fig:dec-blocking-2010}(e,f) and \ref{fig:dec-outside-blocking-2010}(e,f). As for the other two measures, the closeness fields have been averaged over $4$ days and interpolated. Figures~\ref{fig:dec-blocking-2010}(e,f) present the harmonic closeness centrality field on the two selected dates during the 2010 blocking events. The geopotential height contours (white lines) allow us to locate the stationary high pressure cell corresponding to the blocking. We observe that while the pressure cell itself is characterized by very low closeness values, its surroundings are associated with high closeness values. This indicates that the air masses within the blocking structure are isolated and well separated from the main atmospheric currents flowing around it. More precisely, we observe that the high closeness values highlight an atmospheric "track" whose shape resembles that of the eastward flowing subpolar jet stream at $300$~hPa geopotential height level (commonly identified via the maxima of the horizontal wind speed). Hence, as large and nearly stationary meanders of the subpolar jet stream are a known characteristic of blocking situations~\cite{Hakkinen655,Nakamura42}, a manifestation of this feature can be observed in terms of the closeness tracks.

Figures~\ref{fig:dec-outside-blocking-2010}(e,f) show the closeness fields before and after the series of blocking events. Here, no significantly isolated pressure cell is visible in the mid-latitudes and the meridional deviations of the closeness tracks are only minor. In particular we do not observe meandering and circular patterns in the tracks like those occurring during blocking episodes (see Figs.~\ref{fig:dec-blocking-2010}(e,f)).  Studying the full temporal evolution of the closeness field for all three considered years (see videos provided in the Supplementary Material) demonstrates that similar observations apply to the two other years 2003 and 2018 as well.

Despite the overall similarities between the jet pattern and the closeness tracks, some important differences are found as well. First, while the jet is commonly identified at the $300$~hPa pressure level, the closeness is not associated with any particular height. In fact, recall from Section~\ref{sec:computational-aspects} that, after being initialized at $300$~hPa, the particles' trajectories are fully three-dimensional, and that the vertical component of the final position is ignored for the network computation. Nonetheless, as mentioned in Section~\ref{sec:computational-aspects}, we do not expect the two-dimensional projection to produce large differences in the results with respect to a fully three-dimensional partition of the atmosphere.

Second, while the maxima in the wind velocities, identifying the jet streaks, are local properties, the tracks of out-closeness highlight "chains" of regions with strong and non-divergent (i.e., "channeled") wind in the downstream direction (see Section~\ref{subsec:closeness-theory} for further details). As a result, the closeness values tend to be higher at the entrance of jet streaks than at their exit. Besides, while the strongest wind speeds are often similar both upstream and downstream of pressure cells, the closeness values along the tracks are generally comparatively lower upstream of the pressure perturbations than downstream of it. This is well observable in Figs.~\ref{fig:dec-blocking-2010}(f).
In addition, the non-local or multi-step character of the closeness renders the closeness tracks to be continuous, whereas the jet typically consists of a sequence of different separated streaks at the time scale of days.

Recall that, each closeness field has been computed here for a single Lagrangian network involving trajectories of duration $\tau$. A more faithful description of the transport process would require \cite{SerGiacomi2015-blocking,Ser-GiacomiPRE} the use of a sequence of Lagrangian flow networks computed at successive intervals of duration $\tau$ to compute the actual shortest path between two nodes $i$ and $j$ that defines the distance $d(i,j)$.  We have verified, however, that the actual trajectories of tracers closely follow the closeness tracks during times corresponding to a few $\tau$-intervals (see Supplementary Material), thus confirming that the calculation of closeness fields based on single "snapshot" networks provides a useful tool, which is computationally much less demanding than the use of multi-time sequences of temporal networks, \cite{SerGiacomi2015-blocking,Ser-GiacomiPRE} to characterize currents of air mass transport in the atmosphere.

\section{\label{sec:conclusion} Conclusions}

We have constructed Lagrangian flow network representations of the Northern Hemisphere atmospheric circulation from available reanalysis data . Focusing on Northern Hemisphere summer blocking situations, we have characterized the behaviour of a selection of network measures before, during and after large-scale blocking events in three different years (2003, 2010 and 2018). The network patterns highlight (to different degrees) anomalous atmospheric circulation patterns, thereby allowing us to trace the spatio-temporal characteristics of atmospheric blocking events.

More specifically, the per-node degree and entropy, which are both local measures quantifying flux exchange between each region (node) and the rest of the atmosphere occurring during one time step, consistently exhibit low values within the blocking region and high values in its surroundings. This reflects the isolated character of the air mass inside the blocking high pressure cell at the considered time scale (here 24 hours).

Conversely, the harmonic closeness centrality provides a more global point of view by highlighting topological pathways along which material flow is important. Although, in principle, these pathways may not correspond to any real trajectory, in practice, they do agree with the main currents of material transport in the atmosphere. This demonstrates a certain persistence in these currents over the chosen integration time (of 8 hours). In particular, the shape of the closeness tracks shows strong similarities with the subpolar jet stream, highlighting very clearly the meandering of the jet around blocking highs. Beyond general similarities with the spatial pattern of the jet, some interesting differences exist. These differences include the continuous character of the closeness tracks (whereas the jet is divided into streaks on the time scale of days) and the characteristic patterns of the closeness values along these tracks: the highest closeness values are generally observed at the entrance of the jet streaks while the lowest values occur upstream of the source of perturbations.

In conclusion, it is remarkable that three simple network measures (degree, entropy and harmonic closeness centrality) allow to recover key characteristics of blocking events, including the isolated character of the air mass forming the blocking high pressure cell and the deviation of important atmospheric currents around it.

It is worth mentioning again that this work has focused on complex network measures (the degree, entropy and harmonic closeness centrality) characterizing properties of individual nodes. This greatly facilitates the visualization and interpretation of the obtained patterns. Many other methods can be applied to the transport matrices to extract relevant connectivity properties. In particular, spectral methods\cite{vonLuxburg2007} have been extensively used in the literature to find partitions of fluid domains into coherent sets\cite{Froyland2003, Froyland2005, Santitissadeekorn2010}, i.e. regions remaining at the same position in space and relatively non-dispersive over a finite time period. The local quantities described in our work diagnose additional and, in some sense, more subtle aspects of the connectivity. Specifically, as mentioned above, the degree and entropy provide local information about the stretching of the flow: while low values corresponding to isolated flow regions are expected to present similarities with regions identified via spectral partitioning, high values identify regions characterized locally by strong divergence and high velocities, properties which are not highlighted by the spectral methods. Further, while the harmonic closeness centrality clearly distinguishes the pathways of the main currents from the blocking region (highlighted by high and low values respectively), the spectral methods do not provide clear information about the position of such currents. Hence, the local metrics extracted from our Lagrangian Flow Network approach provide valuable additions to already existing tools when studying time-dependent flows. A more thorough comparison and study of the relationship between spectral and node-based approaches will be subject of future studies.

In addition, further developments will be needed to fully exploit the potential of Lagrangian flow networks for the study of atmospheric circulation patterns associated with extreme weather events. In particular, although the harmonic closeness centrality, computed here from single-time networks, proved highly successful in identifying the strongest atmospheric currents, a more systematic comparison with results from pathways in multi-time sequences of networks, representing true trajectories in a statistical sense, would be desirable. Other network measures and the effect of link weighting shall be explored to obtain further insights into the atmospheric transport characteristics. Finally, in order to study a larger variety of atmospheric phenomena the use of three-dimensional Lagrangian Flow Networks will be necessary.

\section*{\label{sec:supplementary} Supplementary Material}
The \href{https://cloud.ifisc.uib-csic.es/nextcloud/index.php/s/eW52QqstqJAsiJ6}{Supplementary Material} accompanying this manuscript contains videos showing the time evolution of the Lagrangian flow network in terms of its degree, entropy and harmonic closeness centrality patterns during Mai--August 2010, June--August 2003 and June--August 2018. In addition, we provide a video showing the time evolution of the position of tracers superimposed on the closeness field to support our corresponding argument presented in Section~\ref{subsec:closeness-results}.

\begin{acknowledgments}
This project has received funding from the European Union's
Horizon 2020 research and innovation programme under the Marie
Sk{\l}odowska-Curie grant agreement No813844. RVD has been
additionally supported by the Federal Ministry for
Education and Research of Germany (BMBF) via the JPI
Climate/JPI Oceans project ROADMAP (grant no. 01LP2002B). CL
and EHG also acknowledge support from MINECO/AEI/FEDER through
the Mar\'{\i}a de Maeztu Program for Units of Excellence in
R\&D (MDM-2017-0711, Spain). We thank Mónica Minjares for help
with blocking detection indices and the TM90 index in particular.
\end{acknowledgments}

\section*{Data Availability}
 The networks' weighted adjacency matrices corresponding to the periods Mai--August 2010, June--August 2003 and June--August 2018, as well as the Python code used to obtain the results presented in this work can be found on \href{https://github.com/noemiee/LFN-blocking.git}{GitHub}, along with detailed explanations of the computational procedures.

\nocite{*}
\bibliography{manuscript}

\end{document}